\documentclass[a4paper,11pt]{article}

\usepackage{jcappub} 
\usepackage{epstopdf}
\usepackage{slashed}

\title{Dark matter kinetic decoupling with a light particle}

\author[a]{Ayuki Kamada}
\author[b]{and Tomo Takahashi}

\affiliation[a]{Institute for Basic Science, Center for Theoretical Physics of the Universe, Daejeon 34051, South Korea}
\affiliation[b]{Department of Physics, Saga University, Saga 840-8502, Japan}

\emailAdd{akamada@ibs.re.kr}
\emailAdd{tomot@cc.saga-u.ac.jp}

\abstract{
We argue that the acoustic damping of the matter power spectrum is not a generic feature of the kinetic decoupling of dark matter, but even the enhancement can be realized depending on the nature of the kinetic decoupling when compared to that in the standard cold dark matter model.
We consider a model that exhibits a {\it sudden} kinetic decoupling and investigate cosmological perturbations in the {\it standard} cosmological background numerically in the model. 
We also give an analytic discussion in a simplified setup.
Our results indicate that the nature of the kinetic decoupling could have a great impact on small scale density perturbations.
}

\arxivnumber{
}

\begin{document}

%
\def\aj{AJ}%
\def\actaa{Acta Astron.}%
\def\araa{ARA\&A}%
\def\apj{ApJ}%
\def\apjl{ApJ}%
\def\apjs{ApJS}%
\def\ao{Appl.~Opt.}%
\def\apss{Ap\&SS}%
\def\aap{A\&A}%
\def\aapr{A\&A~Rev.}%
\def\aaps{A\&AS}%
\def\azh{AZh}%
\def\baas{BAAS}%
\def\bac{Bull. astr. Inst. Czechosl.}%
\def\caa{Chinese Astron. Astrophys.}%
\def\cjaa{Chinese J. Astron. Astrophys.}%
\def\icarus{Icarus}%
\def\jcap{J. Cosmology Astropart. Phys.}%
\def\jrasc{JRASC}%
\def\mnras{MNRAS}%
\def\memras{MmRAS}%
\def\na{New A}%
\def\nar{New A Rev.}%
\def\pasa{PASA}%
\def\pra{Phys.~Rev.~A}%
\def\prb{Phys.~Rev.~B}%
\def\prc{Phys.~Rev.~C}%
\def\prd{Phys.~Rev.~D}%
\def\pre{Phys.~Rev.~E}%
\def\prl{Phys.~Rev.~Lett.}%
\def\pasp{PASP}%
\def\pasj{PASJ}%
\def\qjras{QJRAS}%
\def\rmxaa{Rev. Mexicana Astron. Astrofis.}%
\def\skytel{S\&T}%
\def\solphys{Sol.~Phys.}%
\def\sovast{Soviet~Ast.}%
\def\ssr{Space~Sci.~Rev.}%
\def\zap{ZAp}%
\def\nat{Nature}%
\def\iaucirc{IAU~Circ.}%
\def\aplett{Astrophys.~Lett.}%
\def\apspr{Astrophys.~Space~Phys.~Res.}%
\def\bain{Bull.~Astron.~Inst.~Netherlands}%
\def\fcp{Fund.~Cosmic~Phys.}%
\def\gca{Geochim.~Cosmochim.~Acta}%
\def\grl{Geophys.~Res.~Lett.}%
\def\jcp{J.~Chem.~Phys.}%
\def\jgr{J.~Geophys.~Res.}%
\def\jqsrt{J.~Quant.~Spec.~Radiat.~Transf.}%
\def\memsai{Mem.~Soc.~Astron.~Italiana}%
\def\nphysa{Nucl.~Phys.~A}%
\def\physrep{Phys.~Rep.}%
\def\physscr{Phys.~Scr}%
\def\planss{Planet.~Space~Sci.}%
\def\procspie{Proc.~SPIE}%

\maketitle
\flushbottom

\section{Introduction}
\label{sec:intro}
The nature of dark matter (DM) has not been uncovered regardless of accumulated evidences of its existence  from cosmic to galactic scale structures of the Universe (see Ref.\,\cite{Bertone:2016nfn} for a recent review).
The precise measurements of the cosmic microwave background anisotropies, for example, does not only support its existence but also precisely determines the DM present mass density\,\cite{Ade:2015xua}.
There has been vigorous efforts to identify the nature of DM both in direct and indirect ways.
One of such efforts is seeking imprints of DM interactions with other particles like baryons, photons, and neutrinos on the large scale structure of the Universe\,\cite{Boehm:2000gq, Chen:2002yh, Dubovsky:2003yn, Sigurdson:2004zp, Boehm:2004th, Mangano:2006mp, Serra:2009uu, Wilkinson:2013kia, Dolgov:2013una, Cyr-Racine:2013fsa, Dvorkin:2013cea, Wilkinson:2014ksa, Escudero:2015yka, Ali-Haimoud:2015pwa, Lesgourgues:2015wza, Ko:2016uft, Ko:2016fcd}.

The imprints attract growing interests also in the context of the small scale problems in the standard $\Lambda$CDM model (see Ref\,\cite{DelPopolo:2016emo} for a recent review)
where the dark energy ($\Lambda$) and cold and only gravitationally interacting DM (CDM) are assumed.
The DM interactions result in suppressions of small scale density perturbations, the cutoff scale of which is determined by the horizon scale at the time of the kinetic decoupling\,\cite{Loeb:2005pm, Bertschinger:2006nq}.
The reduced number of subgalactic halos may be concordant with the observed number of the dwarf galaxies, which may be overpredicted in the standard CDM model\,\cite{Boehm:2001hm, Sigurdson:2003vy, Profumo:2004qt, vandenAarssen:2012ag, Aarssen:2012fx, Kamada:2013sh, Boehm:2014vja, Buckley:2014hja, Schewtschenko:2014fca, Cyr-Racine:2015ihg, Vogelsberger:2015gpr, Schewtschenko:2015rno, Binder:2016pnr}.
\footnote{
It has also been reported that cosmological hydrodynamic simulations that incorporate baryonic processes reproduce observed small scale structure in the standard CDM model\,\cite{Sawala:2015cdf}.
}

Though previous studies consider that the kinetic decoupling leads to a damped oscillation (acoustic damping) of the resultant matter power spectrum, the physics of the kinetic decoupling is not fully understood.
For example, it is not clear if the acoustic oscillation is always damped in spite of the nature of the kinetic decoupling.
Paper I\,\cite{Kamada:2016qjo}, where charged massive particles are considered to account for the DM mass density, shows that the resultant matter power spectrum can be enhanced when compared to that in the standard $\Lambda$CDM model, depending on the {\it suddenness} of the kinetic decoupling.
The Coulomb scattering between charged massive particles and baryons become inefficient when $e^{-}e^{+}$ annihilations freeze out.
At that time the momentum transfer rate per Hubble time decreases with the temperature suddenly by the Boltzmann factor in the $e^{\pm}$ number density, which results in the enhancement of the DM density perturbations that enters the horizon around the $e^{-}e^{+}$ annihilation.
We note that the enhancement of the resultant matter power spectrum at small scales can also be realized in models where an early matter domination is considered \cite{Zhang:2015era, Choi:2015yma}. 
However, we do not consider such an early matter dominated era and just assume the {\it standard} thermal history. 
The enhancement mechanism due to the sudden kinetic decoupling, which will be discussed in this paper, is quite different from the one arising in models with an early matter dominated era.

In this paper, following paper I\,\cite{Kamada:2016qjo}, we show that the enhancement ({\it overshooting}) of the resultant matter power spectrum is not limited in the case of charged massive particles, but just the suddenness of the kinetic decoupling is the essence of the overshooting and thus it could be applicable to a broad class of models.
We take a phenomenological DM model with a pseudo scalar coupling to a mediator that also couples to hidden neutrinos as an example.
In some parameter choice, the DM kinetic decoupling happens when the particle with which DM interacts becomes non-relativistic and its number density drops with the Boltzmann factor.
It results in the virtually instantaneous kinetic decoupling, while the usual kinetic decoupling leaves an intermittent collisions.
The resultant matter power spectrum shows oscillating features as usual, but the oscillation amplitude is not damped.

The paper is organized as follows.
In Sec.\,\ref{sec:model}, we describe the model setup and its thermal history considered in this paper.
There the energy transfer rate is introduced and calculated, which plays an important role in the evolution of the DM perturbations.
In Sec.\,\ref{sec:evol}, first we present evolution equations of the DM perturbations.
Then we give some analytic result in a simplified assumption to explain why we expect that the acoustic damping disappears when the kinetic decoupling occurs suddenly.
Finally we numerically follow the co-evolution of the cosmological perturbations of DM, mediators, hidden neutrinos, and the other degrees of freedom to confirm the expectation.
Sec.\,\ref{sec:concl} is devoted to concluding remarks.
Throughout this paper we use the {\it Planck} 2013 cosmological parameters\,\cite{Ade:2013zuv} to be consistent with Ref.\,\cite{Binder:2016pnr}: $\Omega_{b} h^{2} = 0.022068$, $\Omega_{c} h^{2}=0.12029$, $H_{0}=67.11$, $\ln(10^{10}A_{\rm s}) = 3.098$, and $n_{\rm s}=0.9624$.

\section{On-shell mediator}
\label{sec:model}
In Ref.\,\cite{Binder:2016pnr} a general aspect of the DM kinetic decoupling in phenomenological models is studied, where we extend the standard model (SM) with a DM sector, which contains a Dirac DM ($\chi$), a bosonic mediator ($\phi$) and $N_{\nu}/2$ copies of hidden Dirac neutrinos ($\nu$)\,\cite{vandenAarssen:2012ag, Aarssen:2012fx}.
The interactions are given by
\begin{eqnarray}
\label{eq:scalar}
{\mathcal L}_{\rm S} 
&\supset& 
g_{\chi} \bar{\chi} \phi \chi + g_{\nu} \bar{\nu} \phi \nu \,, \\
{\mathcal L}_{\rm V} 
&\supset& 
g_{\chi} \bar{\chi} \gamma^{\mu} \chi \phi_{\mu} + g_{\nu}  \bar{\nu} \gamma^{\mu} \nu \phi_{\mu} \,, \\
\label{eq:pscalar}
{\mathcal L}_{\rm PS} 
&\supset& 
i g_{\chi} \bar{\chi} \phi \gamma^{5} \chi + i g_{\nu} \bar{\nu} \phi \gamma^{5} \nu \,,  \\
{\mathcal L}_{\text{PV}} 
&\supset& 
g_{\chi} \bar{\chi} \gamma^{\mu} \gamma^{5} \chi \phi_{\mu} + g_{\nu} \bar{\nu} \gamma^{\mu} \gamma^{5} \nu \phi_{\mu} \,, 
\end{eqnarray}
for the scalar, vector, pseudo scalar, and pseudo vector mediators, respectively.

We consider the pseudo scalar mediator in this paper since it has a different property from the others.
The spin-averaged invariant amplitude squared for $\chi \nu \to \chi \nu$ scales as $|{\mathcal M}|^{2} \propto (-t)^{2}$ in the cases of the pseudo scalar mediator, while $\overline{|{\mathcal M}|^{2}} \propto (-t) m_{\chi}^{2}$, $E_{\nu}^{2} m_{\chi}^{2}$, $E_{\nu}^{2} m_{\chi}^{2}$ in the case of the scalar, vector, and pseudo vector mediators, respectively.
Here we assume that the DM ($m_{\chi}$) and mediator ($m_{\phi}$) masses are much larger than the hidden neutrino energy ($E_{\nu}$).
The different scaling originates from that in the small momentum transfer limit, where the DM particle is regarded as at rest before and after the collision and thus the expectation value of $\bar{\chi} \gamma^{5} \chi$ vanishes.
It follows that the kinetic decoupling through the pseudo scalar mediator occurs much earlier than those through the others with the model parameters being fixed. 

Let us take a closer look.
The momentum transfer rate, which determines not only the the kinetic decoupling temperature but also the density perturbation evolution around the kinetic decoupling\,\cite{Binder:2016pnr}, is given by
\begin{eqnarray}
\label{eq:gamnonrel}
\gamma 
=
\frac{1}{6 m_{\chi} T}
\sum_{s_{\rm TP}} \int \frac{d^{3} \mathbf{p}_{\rm TP}}{(2\pi)^{3}}
f^{\rm eq}_{\rm TP} 
(1 \mp f^{\rm eq}_{\rm TP})
\int^{0}_{-4\mathbf{p}^{2}_{\rm TP}} dt (-t)
\frac{d\sigma}{dt}v
\,,
\end{eqnarray}
where the subscript (TP) denotes the particle in a thermal bath with which DM elastically interact, $T$ is the temperature of the thermal bath, $t$ is the Mandelstam variable that represents the momentum transfer squared, and $f^{\rm eq} = (\exp\{(-p \cdot u - \mu) / T\} \pm 1)^{-1}$ with $\mu$ and $u$ being respectively the chemical potential and the bulk velocity.
The $\mp$ in $\gamma$ ($\pm$ in $f^{\rm eq}$) is for  bosons and fermions, respectively.
The above formula is valid as long as the DM mass is much larger than the temperature.

The momentum transfer rate per Hubble time for $\chi \nu \to \chi \nu$ through the pseudo scalar mediator is given by
\begin{eqnarray}
\label{eq:gamchinu}
\frac{\gamma}{H} 
\simeq 
2 \, \left( \frac{r}{r_0} \right)^{2}  \left( \frac{N_{\nu}}{6} \right) \left( \frac{\alpha_{\chi}}{4.7 \times 10^{-2}} \right) \left( \frac{\alpha_{\nu}}{10^{-12}} \right) \left( \frac{1 \, {\rm GeV}}{m_{\chi}} \right)^{3} \left( \frac{10 \, {\rm keV}}{m_{\phi}} \right)^{4} \left( \frac{T_{\nu}}{10 \, {\rm keV}} \right)^{6} \,,
\end{eqnarray}
where $\alpha_{\chi} = g_{\chi}^{2} / (4 \pi)$ and $\alpha_{\nu} = g_{\nu}^{2} / (4 \pi)$, and $H^{2} = 4 \pi^3 g_{*} T_{\gamma}^{4} / (45 m_{\rm Pl}^{2})$ with $g_{*}$ being the effective number of massless degrees of freedom for the energy density and $m_{\rm Pl}$ being the Planck mass.
Here note that the dark matter relic abundance through the $\chi {\bar \chi} \to \phi \phi$ annihilation is given by
\begin{eqnarray}
\Omega_{\chi} h^2 \simeq \frac{0.12}{2} \, \left( \frac{r}{r_0} \right) \left( \frac{4.7 \times 10^{-2}}{\alpha_{\chi}} \right)^{2} \left( \frac{m_{\chi}}{1 \, {\rm GeV}} \right)^{2} \left( \frac{x_f}{13.4} \frac{r_0}{r} \frac{1}{10^{2}} \right)^2 \,,
\end{eqnarray}
where $r$ is the ratio of the hidden neutrino temperature to the photon temperature at the time of the kinetic decoupling, i.e., $r = T_\nu / T_\gamma$. 
The ratio for SM neutrinos is denoted by $r_0$ and takes a value of $(4/11)^{1/3}$ at $T_{\gamma} = {\mathcal O} (1)$\, keV.
Here the factor ($1/10^{2}$) in the last parentheses can be achieved with a low reheating temperature of $T_{\rm RH} = m_{\chi} /13.4 (r_{0} / r) (1/10^{2})$.
As we see from Eq.\,(\ref{eq:gamchinu}), depending on the parameters, say, for $\alpha_{\nu} \ll 10^{-12}$, the decoupling of $\chi \nu \to \chi \nu$ occurs before $\phi$ becomes non-relativistic, i.e., the mediator mass becomes larger than the hidden neutrino energy.
\footnote{
In this case Eq.\,(\ref{eq:gamchinu}) is not valid any longer since there we assume that the mediator mass is much larger than the hidden neutrino energy.
Essentially $m_{\phi}^{4}$ in the denominator is replaced by $T_{\nu}^{4}$ in addition to a overall multiplication by a hidden {\it Coulomb logarithm}; $\sim \ln (T_{\nu}^{2} / {\rm max} [ m_{\phi}^{2}, \alpha_{\nu} T_{\nu}^{2} ])$.
}
In the following we focus on such a case.

Even after the $\chi \nu \to \chi \nu$ decoupling, $\phi \leftrightarrow \nu {\bar \nu}$ and $\chi \phi \to \chi \phi$ reactions are still efficient as long as $\phi$ is relativistic.
$\phi$'s are kept in thermal equilibrium with $\nu$'s through the decay and inverse decay, which are as efficient as
\begin{eqnarray}
\frac{m_{\phi}}{4 T_{\nu}} \frac{\Gamma}{H} 
\simeq 
10^{11} \, \left( \frac{r}{r_{0}} \right)^{2} \left( \frac{N_{\nu}}{6} \right) \left( \frac{\alpha_{\nu}}{10^{-12}} \right) \left( \frac{m_{\phi}}{10 \, {\rm keV}} \right)^{2} \left( \frac{10 \, {\rm keV}}{T_{\nu}} \right)^{3} \,,
\end {eqnarray}
where the decay rate at rest is evaluated as
\begin{eqnarray}
\label{dec}
\Gamma &=& \frac{\alpha_{\nu}}{4} N_{\nu} m_{\phi} \,.
\end{eqnarray}
The prefactor $(m_{\phi} / 4 T_{\nu})$ represents the boost factor for a relativistic $\phi$ and the Pauli blocking so that the above estimation gives the lower bound.

Let us discuss $\nu$'s after the $\chi \nu \to \chi \nu$ decoupling.
It should be noted that $\phi \leftrightarrow \nu {\bar \nu}$ reaction is efficient for $\phi$'s, but not for $\nu$'s.
This is because the inverse decay is suppressed by the Boltzmann factor but the number density of $\phi$ itself is small due to the same factor, while the number density of $\nu$ is huge and only a tiny fraction of them experiences the inverse decay.
If $\nu$ does not have any other interaction, it starts to freely stream just after the inverse decay becomes inefficient.
This impacts the evolution of the cosmological perturbations and the resultant matter power spectrum as we see in appendix\,\ref{sec:freestream}.
Let us consider the scalar mediator given in Eq.\,(\ref{eq:scalar}) as well as the pseudo scalar mediator and take $g_{\chi, s}$ tiny and $g_{\nu, s}$ sizable.
Then without changing the dynamics of $\chi$ we can introduce the scatterings between $\nu$'s.
If $m_{\phi, s}$ is larger than $T_{\nu}$ of interest, the effective interaction reads as ${\mathcal L} \supset g_{\nu, s}^{2} / m_{\phi, s}^{2} \bar{\nu} \nu \bar{\nu} \nu$.
A rough estimate finds that the scatterings between $\nu$'s are efficient until $T_{\nu} \sim 1 \, {\rm keV} (m_{\phi, s} / g_{\nu, s} / 10 \, {\rm GeV})^{4/3}$.

\begin{figure}[htb]
\begin{center}
\includegraphics[width=0.75\linewidth]{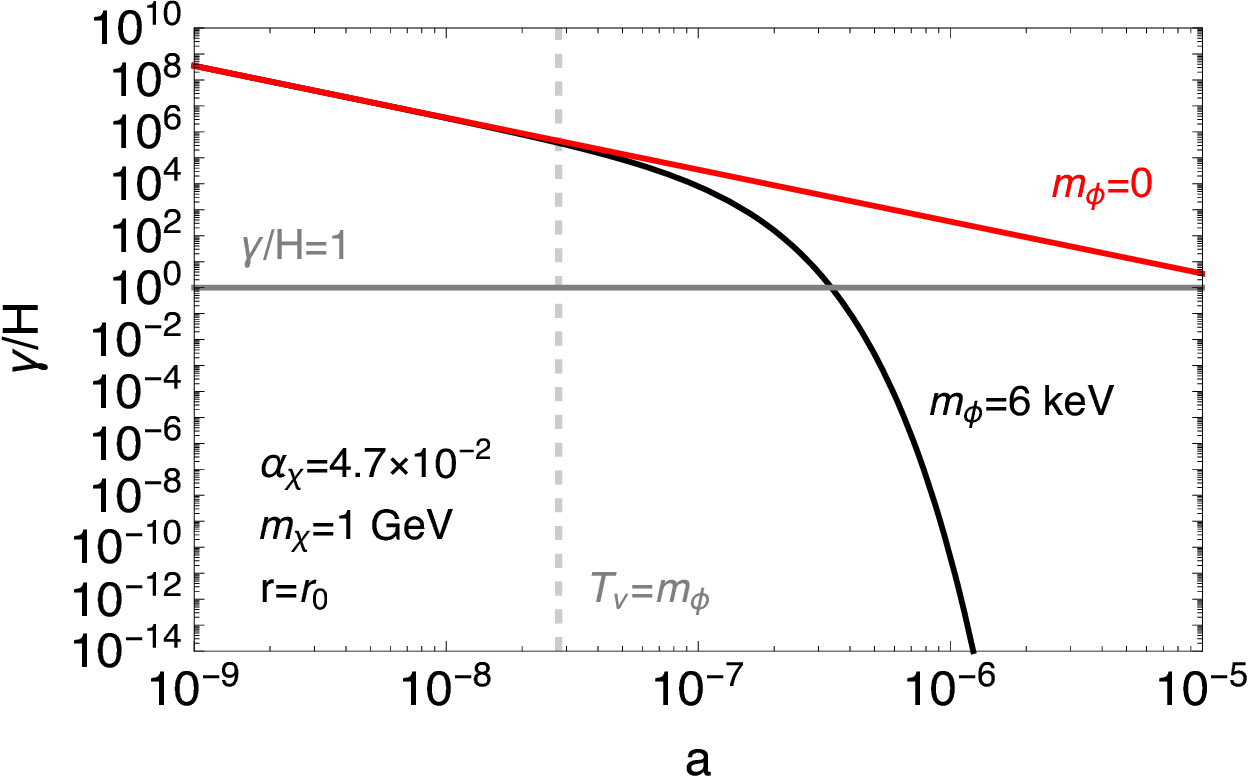}
\caption{
Momentum transfer rate per Hubble time as a function of the scale factor (normalized such that $a = 1$ at present) with $m_{\phi} = 6$\, keV, $\alpha_{\chi} = 4.7 \times 10^{-2}$, $m_{\chi} = 1$\, GeV, and $r = r_{0}$.
The rate starts to deviate from that with a massless mediator ($m_{\phi} = 0$) around $T_{\nu} = m_{\phi}$ (vertical dashed line) and crosses the Gamow criterion ($\gamma/H = 1$, horizontal line) steeply.
}
\label{fig:gamoH}
\end{center}
\end{figure}

\begin{figure}[htb]
\begin{center}
\includegraphics[width=0.75\linewidth]{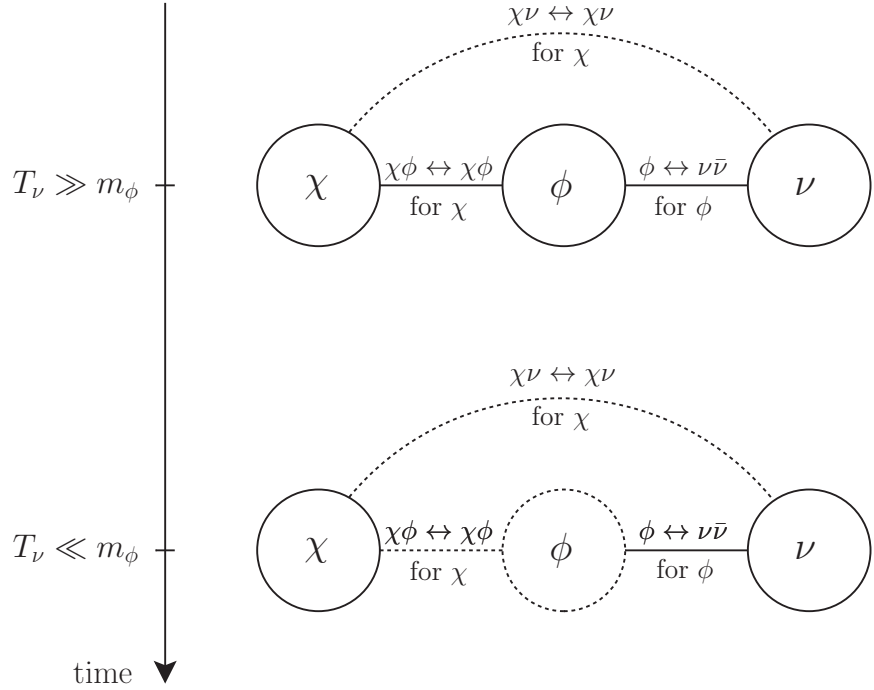}
\caption{Rough sketch of the thermal history of the DM sector.
Dotted lines and circles indicate that the corresponding reactions and particles are decoupled, respectively.
}
\label{fig:history}
\end{center}
\end{figure}

The momentum transfer rate per Hubble time for $\chi \phi \to \chi \phi$ is given by
\begin{eqnarray}
\label{eq:gamchiphi}
\frac{\gamma}{H} 
= 
\frac{4 \langle p_{\phi} \rangle \sigma n_{\phi}}{3 m_{\chi} H}
\stackrel{m_{\phi} \to 0}{\to} 
10^{4} \, \left(\frac{r}{r_0}\right)^2  \left( \frac{\alpha_{\chi}}{4.7 \times 10^{-2}} \right)^{2} \left( \frac{1 \, {\rm GeV}}{m_{\chi}} \right)^{3} \left( \frac{T_{\nu}}{10 \, {\rm keV}} \right)^{2} \,,
\end{eqnarray}
where $\langle p_{\phi} \rangle$ is the thermal average of the absolute value of the three-momentum of $\phi$.
The cross section is constant;
\begin{eqnarray}
\sigma = \frac{2 \pi \alpha_{\chi}^{2}}{m_{\chi}^{2}} \,.
\end{eqnarray}
In the final expression we take the relativistic limit of $\phi$.
Once we take finite $m_{\phi}$ into account, the momentum transfer rate per Hubble time starts to drop around $T_{\nu} = m_{\phi}$ as shown in Fig.\,\ref{fig:gamoH}.
More importantly, the kinetic decoupling happens more suddenly than that in the massless mediator case;
\begin{eqnarray} 
\label{eq:sudcond}
\frac{1}{H^{2}}\frac{d \gamma}{dt}{\Big |}_{\gamma = H} \gg 1 \,,
\end{eqnarray}
which leaves an interesting imprint on the resultant matter power spectrum as we will see∂ in the next section.
Fig.\,\ref{fig:history} sketches the thermal history of the DM sector in this model, which is discussed above.

Before closing this section let us comment on an ultraviolet completion of the DM sector, which we do not specify in this paper.
Since light mediators and hidden neutrinos act as dark radiation, the deviation of the effective number of neutrino species from the standard model value ($\Delta N_{\rm eff}$) can be sizable, which could constrain the model.
Though we do not specify the mechanism that determines $r / r_{0}$ and $T_{\rm RH}$, let us discuss how severe the constraint is.
The $r \left( =T_{\nu}/T_{\gamma} \right)$ should satisfy
\begin{eqnarray}
\frac{r}{r_0} < \left( \frac{\Delta N^{\rm obs}_{\rm eff}}{N_{\nu} + 4 / 7} \right)^{1/4} \,,
\end{eqnarray}
where $\Delta N^{\rm obs}_{\rm eff}$ denots the upper bound on $\Delta N_{\rm eff}$ from observations.
The combined constraint from ``{\it Planck} TT+lowP+lensing+BAO"\,\cite{Ade:2015xua} gives $\Delta N^{\rm obs}_{\rm eff} \simeq 0.65$ (95\%), which implies that $r / r_{0} < 0.56$ for $N_{\nu} = 6$.
The model we consider here can be regarded as an effective theory that describes lower energy physics in the DM sector than the DM mass.
Our aim is studying the kinetic decoupling of DM through the light mediator, which occurs long after the decoupling of heavy particles.
Thus the phenomenological model is sufficient for our purpose.

\section{Evolution of the cosmological perturbations}
\label{sec:evol}
Let us describe the evolution of the cosmological perturbations in the {\it perfect fluid} limit;
the equation of continuity and Euler equation of the DM fluid density perturbation ($\delta_{\chi}$) and the bulk velocity potential ($\theta_{\chi}$) are given by\,\cite{Binder:2016pnr}
\begin{eqnarray}
\label{eq:chipert}
{\dot \delta}_{\chi}
=
-\theta_{\chi}
-\frac{1}{2} {\dot h} \,,
\quad
{\dot \theta}_{\chi}
=
-\frac{\dot a}{a} \theta_{\chi}
+k^{2} c_{\chi}^{2} \delta_{\chi}
+\gamma_{0} a (\theta_{\nu} - \theta_{\chi}) \,,
\end{eqnarray}
with dots denoting the derivatives with respect to the conformal time $\tau$, the subscript  $0$ denoting the quantity at the leading order (i.e., homogeneous and isotropic part), and $k$ being the absolute value of the wavenumber.
The sound speed is defined as
\begin{eqnarray}
c_{\chi}^{2} 
= 
\frac{T_{\chi 0}}{m_{\chi}} \left(1-\frac{1}{3}\frac{d\ln T_{\chi 0}}{d \ln a} \right) \,,
\end{eqnarray}
with the temperature evolving according to
\begin{eqnarray}
\frac{d \ln (a^{2} T_{\chi 0})}{d \tau}
=
2 \gamma_{0} a \left(\frac{T_{\nu 0}}{T_{\chi 0}} -1\right) \,.
\end{eqnarray}
Note that the bulk velocity potential of the $\phi$ fluid is kept the same as that of the $\nu$ fluid since $\phi$'s and $\nu$'s are kept in thermal equilibrium.
This is why $\theta_{\nu}$ appears in the Euler equation though the direct coupling between DM and $\nu$ is not efficient due to the suppression of $\alpha_{\nu}$ as discussed in the previous section.
Hereafter we take the synchronous gauge\,\cite{Ma:1995ey};
\begin{eqnarray}
&&
ds^{2} 
= 
a^{2}(\tau) \{- d\tau^2 + (\delta_{i j} + h_{i j}) dx^i dx^j \} \,, 
\quad
h_{i j} 
=
\frac{\partial_i \partial_j}{\partial^2} h + \left( \frac{\partial_i \partial_j}{\partial^2} - \frac{\delta_{i j}}{3} \right) 6 \eta \,, 
\\
&&
{\ddot h} + \frac{\dot a}{a} {\dot h} 
= 
- 8 \pi G a^{2} \sum (\rho_{0} \delta + 3 \delta P) \,.
\end{eqnarray}

Before following numerically the co-evolution of the cosmological perturbations of DM, mediators, (both SM and hidden) neutrinos, baryons, photons, and the gravitational potentials, let us derive an analytic result in a simplified case, where we ignore the sound speed of DM, i.e., take a heavy DM limit while keeping $\gamma_{0}$ being intact.
In fact, the heavy DM limit is also important when we validate that DM can be described as a perfect fluid since the effects of the imperfectness are suppressed by $k \sqrt{T_{\nu, 0} / m_{\chi}} / (a H)$\,\cite{Binder:2016pnr}.
This point will be examined closely in appendix\,\ref{sec:imperfect}.
Furthermore we assume that  radiation component is dominated by photons and hidden neutrinos, i.e., ignore the contribution from SM neutrinos, which freely stream and cannot be approximated by a perfect fluid.
On the other hand, photons can be regarded as a perfect fluid due to the efficient Compton scattering with baryons.
In addition we introduce the self-interaction between hidden neutrinos to prevent them from freely streaming, which simplifies the equations.

Under these assumptions we obtain a closed set of the equations;
\begin{eqnarray}
{\dot \delta}_{\gamma (\nu)}
=
-\frac{4}{3}\theta_{\gamma (\nu)}
-\frac{2}{3} {\dot h} \,,
\quad
{\dot \theta}_{\gamma (\nu)}
=
\frac{1}{4}k^{2} \delta_{\gamma (\nu)} \,,
\quad
{\ddot h} + \frac{1}{\tau} {\dot h} = - \frac{6}{\tau^{2}} ((1-f_{\nu}) \delta_{\gamma} + f_{\nu} \delta_{\nu})\,
\end{eqnarray}
where $f_{\nu} = \rho_{\nu} / (\rho_{\gamma} + \rho_{\nu})$ is the ratio of the hidden neutrino energy density to the whole radiation (photon + hidden neutrino) energy density.
The solutions are given by\,\cite{Bashinsky:2002vx, Weinberg:2002kg}
\begin{eqnarray}
\label{eq:hdotsol}
{\dot h}
&=&
- \frac{24 C}{\tau} \left( \frac{2}{x} \sin x + \frac{2}{x^{2}} \cos x - \frac{2}{x^{2}} - 1 \right) \,,
\\
\label{eq:deltasol}
\delta_{\gamma}
&=&
\delta_{\nu}
= 
-8 C \left( \frac{2}{x} \sin x - \cos x + \frac{2}{x^{2}} \cos x - \frac{2}{x^{2}} \right) \,,
\\
\label{eq:thetasol}
\theta_{\gamma}
&=&
\theta_{\nu}
=
- \frac{6 C}{\tau} \left( - x \sin x - 2 \cos x + 2 \right) \,,
\end{eqnarray}
where $x = k \tau / \sqrt{3}$.
Here we normalize the perturbations such that
\begin{eqnarray}
{\dot h} 
= 
2 C (k^{2} \tau) \,,
\quad  
\delta_{\gamma}
= 
\delta_{\nu}
=
- \frac{2}{3} C (k \tau)^{2} \,,
\quad
\theta_{\gamma}
=
\theta_{\nu}
=
- \frac{1}{18} C (k^{4} \tau^{3}) \,,
\end{eqnarray}
in the small $x$ limit, which coincide with those for the adiabatic mode given in Eq.\,(96) of Ref.\,\cite{Ma:1995ey}.
Note that $\delta_{\nu} = \delta_{\gamma}$ and $\theta_{\nu} = \theta_{\gamma}$ for the adiabatic mode as long as both $\nu$'s and $\gamma$'s can be regarded to form the perfect fluids, i.e., their free-streaming can be neglected.

We can obtain the DM perturbations by solving
\begin{eqnarray}
{\dot \delta}_{\chi}
=
-\theta_{\chi}
-\frac{1}{2} {\dot h} \,,
\quad
{\dot \theta}_{\chi}
=
-\frac{\dot a}{a} \theta_{\chi}
+\gamma_{0} a (\theta_{\nu} - \theta_{\chi}) \,,
\end{eqnarray}
where Eqs.\,(\ref{eq:hdotsol}) and (\ref{eq:thetasol}) are substituted.
We can make an ansatz for the solutions as\,\cite{Bertschinger:2006nq}
\begin{eqnarray}
\delta_{\chi}
&=&
- 12 C \left( - {\rm Ci} (x) + \frac{1}{x} \sin x + \frac{1}{x^{2}} \cos x - \frac{1}{x^{2}} + f_{1} \ln x + f_{2} \right) \,,
\\
\theta_{\chi}
&=& \frac{12 C}{\tau} \left( f_{1} - 1 \right) \,,
\end{eqnarray}
where the cosine integral follows
\begin{eqnarray}
{\rm Ci} (x)
=
\gamma_{\rm E} + \ln x + \int^{x}_{0} \frac{\cos t - 1}{t} dt \,,
\quad
\frac{d}{d x}{\rm Ci} (x)
= \frac{\cos x}{x} \,,
\end{eqnarray}
with the Euler's constant being $\gamma_{\rm E} \simeq 0.577$.
We end up with a simple set of equations;
\begin{eqnarray}
\label{eq:f1f2}
{\dot f}_{1} + \gamma_{0} a f_{1} 
= 
\gamma_{0} a \left( \cos x + \frac{1}{2} x \sin x \right) \,,
\quad
{\dot f}_{2} + {\dot f}_{1} \ln x 
= 
0 \,.
\end{eqnarray}

In the pure CDM limit, where $\gamma_{0} = 0$, we can set $f_{1} = 1$ and $f_{2} = \gamma_{\rm E} - 1/2$, which satisfies Eq.\,(\ref{eq:f1f2}) and the initial condition of $\delta_{\chi} = -1/2 C (k \tau)^{2}$ for $x \to 0$ and $\theta_{\chi} = 0$, coinciding with that for the adiabatic mode given in Eq.\,(96) of Ref.\,\cite{Ma:1995ey}. 
Note that in this limit the above solution reproduces the well-known logarithmic growth of the CDM density perturbations, $\delta \propto \ln a$, inside the horizon.
Supposing that $\gamma_{0} a \tau \to \infty$ for $\tau \to 0$, we obtain the formal solutions of Eq.\,(\ref{eq:f1f2});
\begin{eqnarray}
f_{1}
&=& 
\int^{\tau}_{0} \frac{d\tau'}{\tau'} \exp \left( - \int^{\tau}_{\tau'} \frac{d\tau''}{\tau''} \gamma_{0}'' a'' \tau'' \right) \gamma_{0}' a' \tau' \left( \cos x' + \frac{1}{2} x' \sin x' \right) \notag \\
\label{eq:f1int}
&=&
\cos x + \frac{1}{2} x \sin x - \int^{\tau}_{0} \frac{d\tau'}{\tau'} \exp \left( - \int^{\tau}_{\tau'} \frac{d\tau''}{\tau''} \gamma_{0}'' a'' \tau'' \right) \left( - \frac{1}{2} x' \sin x' + \frac{1}{2} x'^{2} \cos x' \right),  \notag \\
 \\
\label{eq:f2int}
f_{2}
&=&
\int^{\tau}_{0} \frac{d \tau'}{\tau'} \left( f_{1}' - 1 \right) + \left( 1 - f_{1} \right) \ln x + \gamma - \frac{1}{2}
 \,,
\end{eqnarray}
where (double) primes denote quantities evaluated at $\tau'$ $(\tau'')$. 
The initial conditions of the formal solutions take a form of
\begin{eqnarray}
f_{1} 
= 
\cos x + \frac{1}{2} x \sin x \,,
\quad
f_{2}
=
{\rm Ci}(x) - f_{1} \ln x - \frac{1}{2} \cos x \,,
\end{eqnarray}
for $\tau \to 0$, 
which result in
\begin{eqnarray}
\delta_{\chi} 
= 
\frac{3}{4} \delta_{\nu}
= 
\frac{3}{4} \delta_{\gamma} \,, 
\quad
\theta_{\chi}
=
\theta_{\nu}
=
\theta_{\gamma} \,.
\end{eqnarray}
This initial condition for interacting DM coincides with that discussed in paper I\,\cite{Kamada:2016qjo}, where the residual gauge degrees of freedom in the synchronous gauge is also examined.
Again note that we consider the adiabatic mode and the second equalities are held in the perfect fluid limit of $\nu$ and $\gamma$.

We are interested in the resultant late time density perturbations and thus let us take $\tau \to \infty$ in Eqs.\,(\ref{eq:f1int}) and (\ref{eq:f2int}).
Then $f_{1}$ and $f_{2}$ become functions of $k$.
In general we need to rely on numerical methods to evaluate the integrals in Eqs.\,(\ref{eq:f1int}) and (\ref{eq:f2int}).
On the other hand, if the momentum transfer rate per Hubble time follows a simple scaling law, say,
\begin{eqnarray}
\label{eq:gamma_a_tau}
\gamma_{0} a \tau = \left( \frac{\tau_{d}}{\tau} \right)^{n+4} \,,
\end{eqnarray} 
we can analytically evaluate the integrals with the help of the steepest descent method as done in Ref.\,\cite{Bertschinger:2006nq}.
There $n$ is set to be $0$, but we can generalize the result to an arbitrary value of $n$ such that
\begin{eqnarray}
f_{1} 
\simeq 
\left( \frac{4\pi}{5+n} q \right)^{1/2} \exp \left( - R \right) \left[ \cos \left( I - \frac{4+n}{5+n} \frac{\pi}{4} \right) + q \cos \left( I - \frac{4+n}{5+n} \frac{3\pi}{4} \right) \right] \,, 
\end{eqnarray}
where $R$ and $I$ are respectively defined as the real and imaginary parts of
\begin{eqnarray}
R + i I
=
2  \frac{5+n}{4+n} q \exp \left( i  \frac{4+n}{5+n} \frac{\pi}{2} \right) \,,
\end{eqnarray}
and
\begin{eqnarray}
q
=
\frac{1}{2} \left( \frac{k \tau_{d}}{\sqrt{3}} \right)^{(4+n)/(5+n)} \,.
\end{eqnarray}
The factor of $\exp \left( - R \right)$ represents the damping of the acoustic oscillations for the density perturbations that are subhorizon around the kinetic decoupling ($k > 1/\tau_{d}$).
This damping is often called the acoustic damping in the literature and originates from the intermittent collisions (collision period longer than that of the oscillation) around the kinetic decoupling.
Note that the acoustic damping is different from the diffusion ({\it Silk}) damping and the collisionless (free-streaming) damping\,\cite{Loeb:2005pm}.
In the large $n$ limit (effectively $n = 10 \text{--} 20$ in Fig.\,\ref{fig:gamoH}), which corresponds to the sudden kinetic decoupling,
\begin{eqnarray}
\label{eq:R_and_I}
R
\approx
\frac{\pi}{4+n} \frac{k \tau_{d}}{2 \sqrt{3}} \,,
\quad
I
\approx
\frac{5+n}{4+n} \frac{k \tau_{d}}{\sqrt{3}} \,.
\end{eqnarray}
Interestingly, the damping scale, where $R$ becomes of order of unity, is proportional to $n$ and thus shifted to a smaller length scale with $n$ increasing.
From this observation we expect that the sudden kinetic decoupling leaves the resultant matter power spectrum oscillating but not damped.
\footnote{
Overshooting of the matter power spectrum due to the sudden kinetic decoupling was first pointed out in paper I\,\cite{Kamada:2016qjo}. 
This phenomena has also been observed in Ref.\,\cite{Sarkar:2017vls} where density perturbations in a model with long-lived charged massive particles have been investigated. 
In the paper, the coupling between charged massive particles and baryons has been abruptly switched off 
at the lifetime of the charged massive particle, which is virtually the same as the sudden kinetic decoupling discussed here.
We thank the authors of Ref.\,\cite{Sarkar:2017vls} for the clarification of their calculations\,\cite{private_comm}.
}

It should also be noted here that $R$ roughly corresponds to the ratio of the time scale of the sudden kinetic decoupling to that of the acoustic oscillation.
When the kinetic decoupling proceeds more rapidly than the oscillation, i.e., 
\begin{eqnarray} 
\label{eq:sudcondp}
\frac{1}{\gamma_{0} / H} \frac{d (\gamma_{0} /H )}{dt}{\Big |}_{\gamma_{0} = H} > \frac{k c_{\rm s}}{a} \,,
\end{eqnarray}
with the sound speed of the DM-neutrino fluid being $c_{\rm s}$, the acoustic damping does not occur.
In fact $R$ can be represented as the ratio of the left-handed side to the right-handed side of Eq.\,(\ref{eq:sudcondp});
\begin{equation}
R \sim  \left( \frac{k c_{\rm s}}{a} \right) {\Big /} \left( \frac{1}{\gamma_{0} / H} \frac{d (\gamma_{0} /H )}{dt}{\Big |}_{\gamma_{0} = H} \right).
\end{equation}
By using Eq.\,(\ref{eq:gamma_a_tau}) one can find that this expression also gives almost the same estimate as given in Eq.\,(\ref{eq:R_and_I}).

\begin{figure}[htb]
\begin{center}
\includegraphics[width=0.75\linewidth]{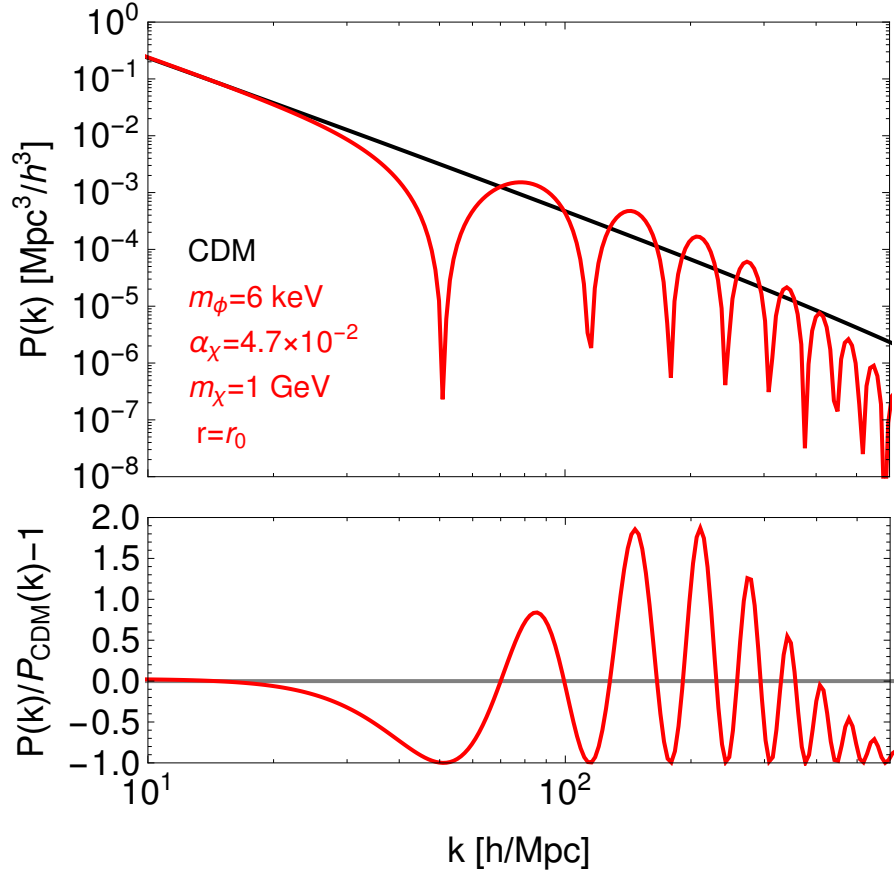}
\caption{Linear matter power spectra extrapolated to the present ($z=0$).
We compare those in the standard CDM model and the interacting DM model with the same parameter choice as in Fig.\,\ref{fig:gamoH}.
The latter clearly shows the oscillation featuring the acoustic oscillation.
Interestingly the powers at the peaks are larger when compared to the standard CDM model. 
}
\label{fig:linear_pf_vb}
\end{center}
\end{figure}

Let us confirm the above expectation by numerically solving the co-evolution of the cosmological perturbations of DM, mediators, (both SM and hidden) neutrinos, baryons, photons, and the gravitational potentials.
To this end we modify the public code \verb|CAMB| suitably\,\cite{Lewis:1999bs}.
Here we take account of the finite sound speed of the DM fluid and assume that the scatterings between $\nu$'s are efficient (as discussed in Sec.\,\ref{sec:model} and see also appendix\,\ref{sec:freestream}).
The resultant linear matter power spectrum is shown in Fig.\,\ref{fig:linear_pf_vb}.
The powers at the oscillation peaks characteristic of the dark acoustic oscillation in interacting DM models are not only undamped as expected, but in fact enhanced when compared to those in the standard CDM model.

\begin{figure}[htb]
\begin{center}
\includegraphics[width=0.75\linewidth]{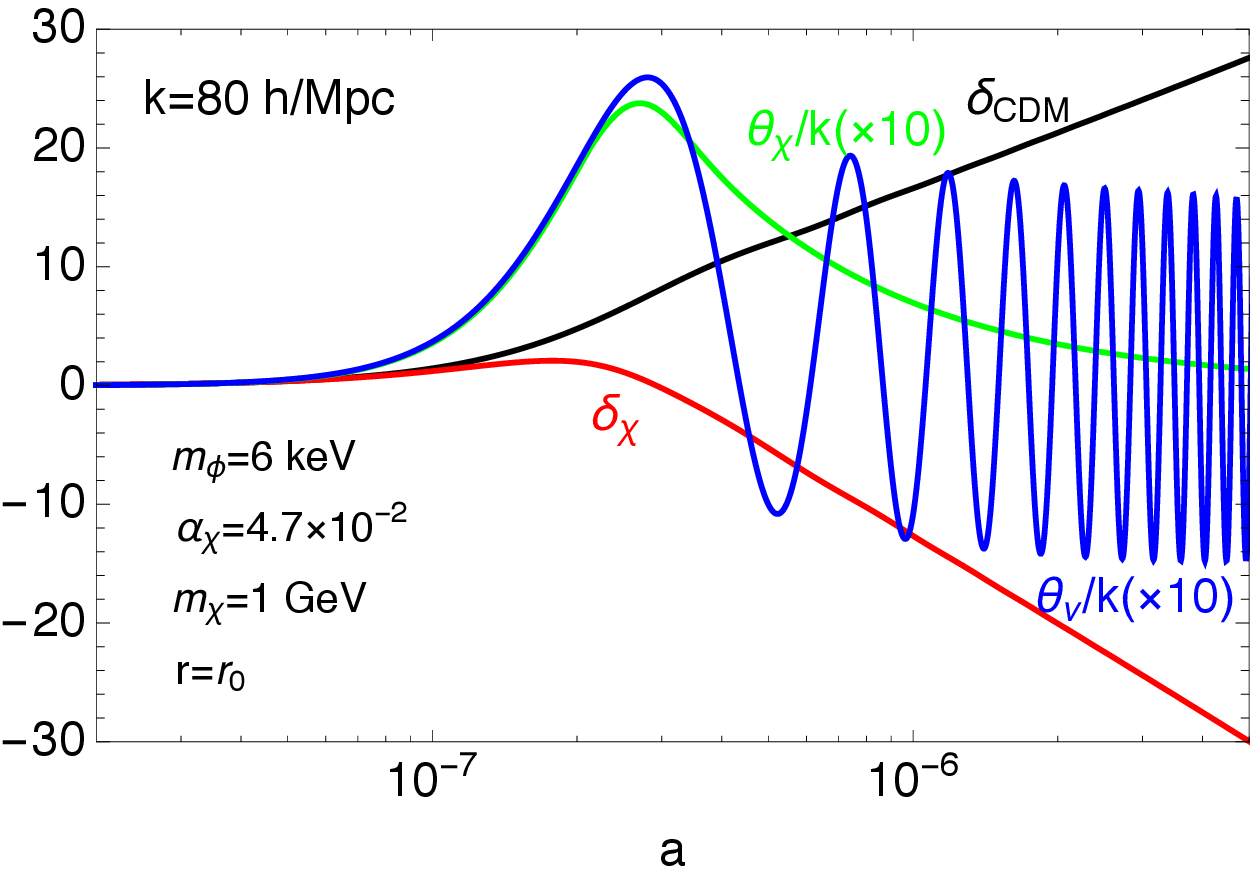} 

\includegraphics[width=0.75\linewidth]{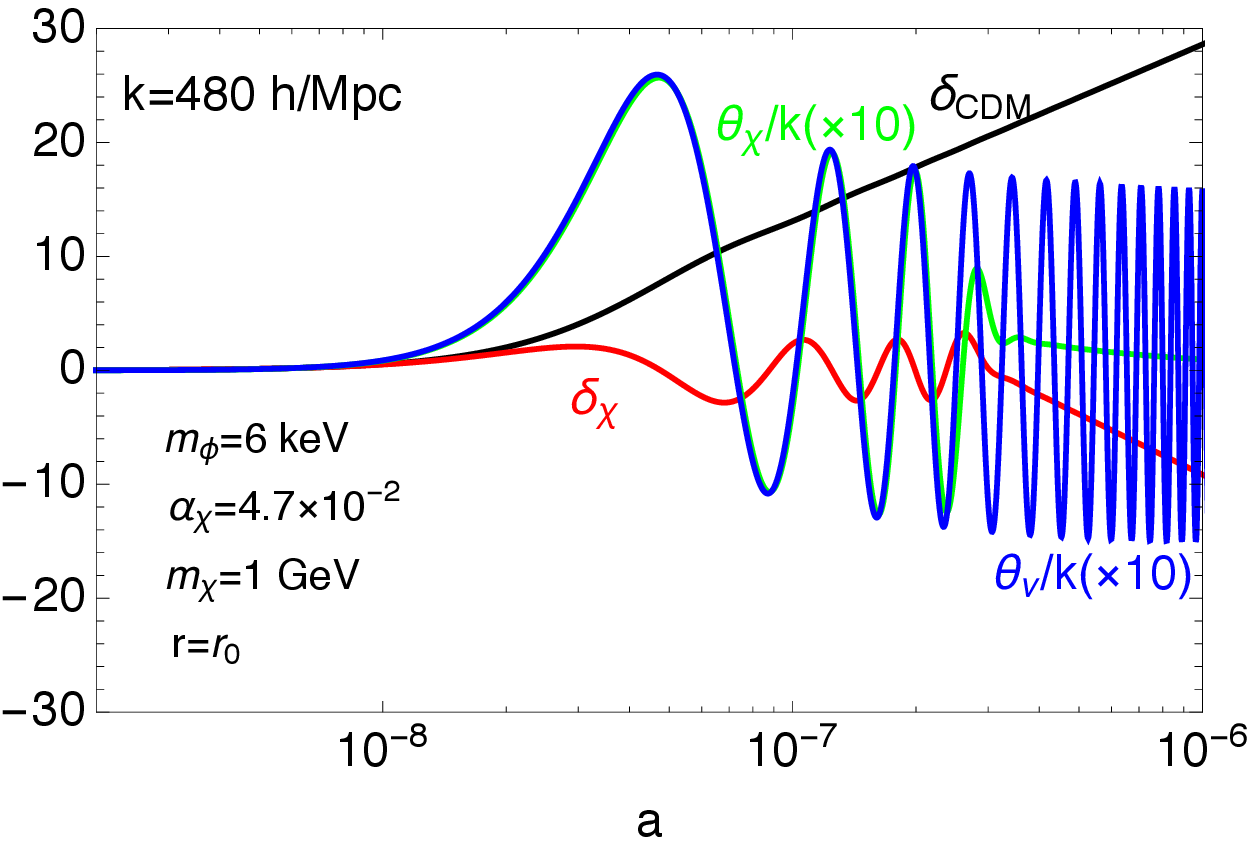}
\caption{
Time evolution of the perturbations with $k = 80 \, h/$Mpc (top) and $480 \, h/$Mpc (bottom).
We multiply the bulk velocity potential of the DM ($\theta_{\chi}$) and neutrino ($\theta_{\nu}$) fluids by a factor of $10$ just for presentation.
We set the parameters of the interacting dark matter model to be the same as in Fig.\,\ref{fig:gamoH}.
}
\label{fig:evol}
\end{center}
\end{figure}

Fig.\,\ref{fig:evol} is helpful to understand the origin of the overshooting.
Here we show the evolution of perturbations with $k = 80$  and $480 \, h/$Mpc.
Initially $\delta_{\chi}$ and $\delta_{\rm CDM}$ coincide with each other, but for $k=80 \, h/$Mpc they end up with the different sign and the absolute value of $\delta_{\chi}$ is larger than $\delta_{\rm CDM}$.
This can be understood as follows.
The bulk velocity potential of the $\chi$ fluid follows that of the $\nu$ fluid due to the $\chi \nu \to \chi \nu$ scattering until the kinetic decoupling.
This drives the oscillation of $\delta_{\chi}$ as well.
If $\chi$'s decouple from $\nu$'s suddenly when $\delta_{\chi} \simeq 0$ and $\theta_{\chi} (= \theta_{\nu})$ is at around the peak, which is in fact the case with $k = 80 \, h/$Mpc as can be seen in the top panel of Fig.\,\ref{fig:evol}, $\theta_{\chi}$ starts to decay as $\theta_{\chi} \propto 1/a$ but drives $\delta_{\chi}$ to the opposite sign before it completely decays.
Intuitively this can be understood by the analogy to the harmonic oscillator. 
In the tight-coupling limit, i.e., before the kinetic decoupling, the evolution of $\delta_{\chi} = (3/4) \delta_{\nu}$ and $\theta_{\chi} = \theta_{\nu}$ can be regarded as a particle oscillating in the quadratic potential. 
The DM density perturbation and the pressure from the $\nu$'s correspond to a position of the particle and the quadratic potential for it, respectively.
Suppose that the potential disappears suddenly at some time, which is an analogy to the sudden kinetic decoupling.
The behavior of the particle after that depends on its position at the time of the potential disappearance.
If it is at the peak, it does not move after that; meanwhile if it is at the bottom of the potential, it starts to stream freely (in reality the bulk velocity potential redshifts), which results in the overshooting of the particle.
The perturbations with $k = 80 \, h/$Mpc correspond to the case where the potential disappears when the particle comes to the bottom of the potential, which is the reason why $\delta_\chi$ is overshooting.

In contrast, for the perturbations with $k=480 \, h/$Mpc depicted in the bottom panel of Fig.\,\ref{fig:evol}, the kinetic decoupling does not occur suddenly when compared to the oscillation time scale (wavenumber times the sound speed). 
$\theta_{\chi}$ starts to deviate from $\theta_{\nu}$ when $\delta_\chi$ takes a value of zero, i.e., when the particle comes to the bottom of the potential in the analogy discussed above. 
However, the intermittent collisions after that prevent $\theta_{\chi}$ with $k=480 \, h$/Mpc from freely decaying ($\theta_{\chi} \propto 1/a$), unlike for $k=80 \, h$/Mpc.
Therefore, in this case, no overshooting occurs and the density perturbation damps as in the usual case.

Since the overshooting is realized when density perturbations do not experience the acoustic damping, 
the condition of overshooting is given by Eq.\,(\ref{eq:sudcondp}).
Therefore the density perturbation with a smaller wavenumber exhibits overshooting in Fig.\,\ref{fig:evol}, but that with a larger wavenumber does not.
Note that Eq.\,(\ref{eq:sudcond}) roughly corresponds to the overshooting of the first peak of the resultant matter power spectrum since $k c_{\rm s} / a \simeq H$ when $\gamma = H$ as long as the entropy density of the fluid is dominated by the radiation degrees of freedom (neutrino in the present case).
The above argument is concordant with that of paper I\,\cite{Kamada:2016qjo}.

Before closing this section, we comment on the effect of the gravitational potential on the above arguments.
Precisely speaking, the gravitational potential ${\dot h}$ also impacts the evolution of $\delta_{\chi}$ (see Eq.\,(\ref{eq:chipert})), but it is subdominant.
This is because both ${\dot h}$ (see Eq.\,(\ref{eq:hdotsol})) and $\theta_{\chi}$ decays inversely proportionally to $a$ when subhorizon and after the kinetic decoupling, but the latter ($\theta_{\nu} = \theta_{\gamma} \propto x_{\rm kd} / \tau_{\rm kd}$) dominates the former (${\dot h} \propto 1 / \tau_{\rm kd}$) for perturbations in which $\theta_{\chi}$ takes a peak value at the time of the kinetic decoupling ($\delta_{\gamma} (x_{\rm kd}) = 0$; $x_{\rm kd} \simeq 4$ for the first peak of the matter power spectrum) as we can see from Eqs.\,(\ref{eq:hdotsol})-(\ref{eq:thetasol}).

\section{Conclusion}
\label{sec:concl}
We studied the impact of the suddenness of the DM kinetic decoupling on the evolution of the DM perturbations and thus the resultant matter power spectrum.
We took a phenomenological model with the pseudo scalar mediator coupling to DM and hidden neutrinos as an example.
If the mediator-neutrino coupling is suppressed, the DM-neutrino scattering become inefficient in the early Universe, but they couple with each other indirectly through the DM-mediator scattering and DM-neutrino decay and inverse decay.
Then the DM kinetic decoupling proceeds when the mediator becomes non-relativistic.
The momentum transfer rate per Hubble time suddenly drops as the number density of the mediator is reduced by the Boltzmann factor.

We numerically followed the co-evolution of cosmological perturbations of DM, mediators, neutrinos, and the other degrees of freedom.
Our calculation relies on the formulation developed in Ref.\,\cite{Binder:2016pnr}.
The resultant matter power spectrum features the dark acoustic oscillation and the powers of the peaks are not damped but enhanced when compared to the standard CDM model.
It will be interesting to study the non-linear matter distribution of the Universe in this type of model.
Our result supports a conjecture in paper I\,\cite{Kamada:2016qjo} that the sudden kinetic decoupling leads to an overshooting of the matter power spectrum and this phenomena can happen in a broad class of models where the suddenness of the kinetic decoupling is satisfied. 
We also analytically studied the evolution of the DM perturbations in a simplified setup to provide a further support to the conjecture.
We note that the overshooting is different from the enhancement that can occur in a non-standard background 
evolution like an early matter dominated era, in which new features of the resultant matter power spectrum have been reported\,\cite{Zhang:2015era, Choi:2015yma} and expected\,\cite{Waldstein:2016blt} recently.

Our result shows that the kinetic decoupling still needs to be more clarified even if we assume the standard background evolution of the Universe.
In this paper we focused on the case where the dark acoustic oscillation can be seen at subgalactic scales, which could attract interest in the context of the issues reported in the structure formation in the standard CDM model.
Even if the kinetic decoupling occurs in the earlier stage of the Universe and thus the resultant dark acoustic oscillation is found at much smaller scales, the kinetic decoupling may impact the minimal halo mass of the Universe to leave the observational effect, say, through the indirect detection signal of DM.
It follows that the vigorous efforts into the detailed understanding of the physics of the DM kinetic decoupling are essential for the identification of the nature of DM.

\acknowledgments
We thank Toyokazu Sekiguchi for carefully reading the manuscript and giving a fruitful feedback.
This work was partially supported by IBS under the project code IBS-R018-D1 (AK), JSPS KAKENHI Grant Number 15K05084 (TT), and MEXT KAKENHI Grant Number 15H05888 (TT).

\appendix

\section{Effects of the imperfectness of the DM fluid}
\label{sec:imperfect}
As stressed in Sec.\,\ref{sec:evol}, the acoustic damping discussed in this paper has a different physical origin from the diffusion damping and the collisionless damping.
The perfect fluid approximation, where the collisionless damping is neglected, gives an ideal setup for studying the acoustic damping.
One may, on the other hand, would like to see what happens in a realistic model, where all such mechanisms may play equally important roles. 
More specifically, in the phenomenological model we take in this paper (see Sec.\,\ref{sec:model}), the dark matter mass cannot be arbitrarily large; otherwise the $\chi \phi \to \chi \phi$ scattering decouples too early to address the small-scale issues (see Eq.\,\ref{eq:gamchiphi}).
In this section we examine to what extent the free-streaming of DM, i.e., imperfectness of the DM fluid impacts the resultant matter power spectrum.  

To this end, in principle, we need to follow the full Boltzmann hierarchy somehow\,\cite{Bertschinger:2006nq, Binder:2016pnr}, while it is beyond the scope of this paper.
Instead we rely on the {\it imperfect fluid} approximation, where the next leading order contributions in terms of $k \sqrt{T_{\nu, 0} / m_{\chi}} / (a H)$ are incorporated\,\cite{Binder:2016pnr}.
There we introduce the anisotropic inertia ($\sigma_{\chi}$) and the entropy perturbation ($\pi_{\chi}$) of the DM fluid in addition to the density perturbation and the bulk velocity potential.
They evolve with the conformal time as follows;
\begin{eqnarray}
\label{eq:chipertdelta}
&&
{\dot \delta}_{\chi}
=
-\theta_{\chi}
-\frac{1}{2} {\dot h} \,,
\\
&&
\label{eq:chiperttheta}
{\dot \theta}_{\chi}
=
-\frac{{\dot a}}{a} \theta_{\chi}
-k^{2} \sigma_{\chi}
+k^{2} (c_{\chi}^{2} \delta_{\chi} + \pi_{\chi})
+\gamma_{0} a (\theta_{\nu} - \theta_{\chi}) \,,
\\
&&
\label{eq:chipertsigma}
{\dot \sigma}_{\chi}
=
-2 \frac{{\dot a}}{a} \sigma_{\chi}
+\frac{4}{3} \frac{T_{\nu 0}}{m_{\chi}} \theta_{\chi} \,
+\frac{2}{3} \frac{T_{\nu 0}}{m_{\chi}} ({\dot h} + 6 {\dot \eta})
-2 \gamma_{0} a \sigma_{\chi} \,,
\\
&&
\label{eq:chipertpi}
{\dot \pi}_{\chi}
=
-2 \frac{{\dot a }}{a} \pi_{\chi}
-\frac{1}{a^{2}} \frac{d (a^{2} c_{\chi}^{2})}{d \tau} \delta_{\chi}
-\left (\frac{5}{3}\frac{T_{\nu 0}}{m_{\chi}} - c_{\chi}^{2}\right) \theta_{\chi}
-\frac{1}{2} \left (\frac{5}{3}\frac{T_{\chi 0}}{m_{\chi}} - c_{\chi}^{2}\right) {\dot h}
\notag \\
&& \qquad
-2 \gamma_{0} a \left[ \pi_{\chi} - \frac{\delta T_{\nu}}{m_{\chi}} - \left( \frac{T_{0}}{m_{\chi}} - c_{\chi}^{2} \right) \delta_{\chi} \right]
+2 \gamma_{0} a \left( \frac{T_{\nu 0}}{T_{\chi 0}} - 1 \right) \frac{T_{\chi 0}}{m_{\chi}} \frac{\delta \gamma}{\gamma_{0}}\,.
\end{eqnarray}
In the present model, the momentum transfer rate is a function solely of the neutrino temperature since $\phi$'s are in thermal equilibrium with $\nu$'s.
Thus we can relate their perturbations such that
\begin{eqnarray}
\frac{\delta \gamma}{\gamma_{0}} = \frac{d \ln \gamma_{0}}{d \ln T_{\nu 0}} \frac{\delta T_{\nu}}{T_{\nu 0}} \,.
\end{eqnarray}
The perturbation of the neutrino temperature is determined by the density perturbation of the neutrino;
$\delta T_{\nu} / T_{\nu 0} = \delta_{\nu} / 4$.

\begin{figure}[htb]
\begin{center}
\includegraphics[width=0.75\linewidth]{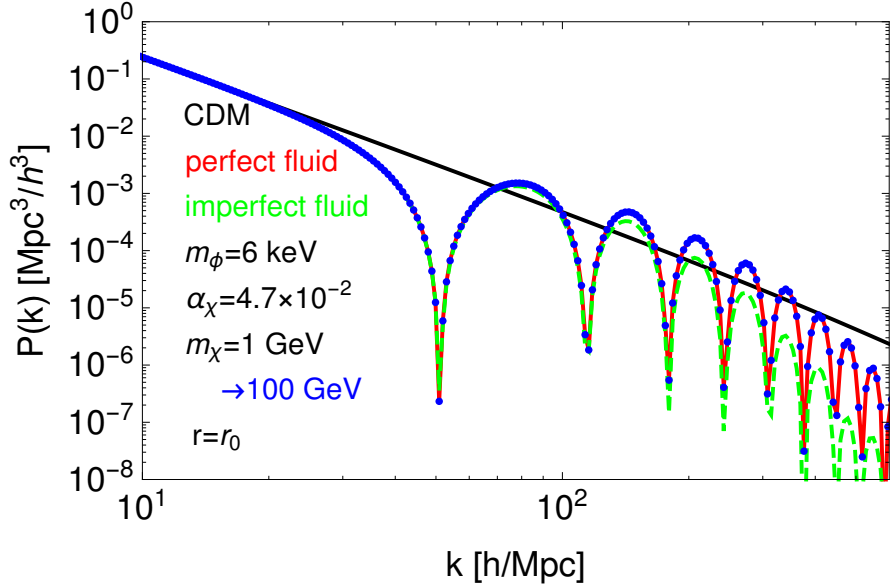}
\caption{Linear matter power spectrum in the imperfect fluid (dashed) is overplotted in Fig.\,\ref{fig:linear_pf_vb}.
They differ from each other above $k \sim 100 \, h/$Mpc.
The first peak ($k \simeq 80 \, h/$Mpc) is not impacted significantly ($10 \%$ at most).
We also show the power spectrum with a DM mass being larger ($m_{\chi} = 1 \to 100$\, GeV) but $\gamma_{0}$ being fixed (dotted).
}
\label{fig:linear_ipf_vb}
\end{center}
\end{figure}

We solve the above equations numerically by suitably modifying \verb|CAMB|\,\cite{Lewis:1999bs}.
Fig.\,\ref{fig:linear_ipf_vb} compares the linear matter power spectrum in the imperfect fluid approximation with that in the perfect fluid approximation shown in Fig.\,\ref{fig:linear_pf_vb}.
The first peak is changed slightly up to 10\% level, while higher peaks changed more drastically.
We need to take into account the full Boltzmann hierarchy, i.e., the effects of the free-streaming carefully at smaller length scales.
The power spectra in the two approximations coincide with each other at large length scales ($k \lesssim 100 \, h/$Mpc), which seems concordant with the rough estimate given in Ref.\,\cite{Binder:2016pnr}; 
the effects of the higher order terms of the Boltzmann hierarchy on the density perturbations are suppressed by $k \sqrt{T_{\nu, 0} / m_{\chi}} / (a H)$, which takes a maximum value at the matter radiation equality.
This ratio is larger than unity for $k > 430 \, h/{\rm Mpc} \, (m_{\chi} / 1 \, {\rm GeV})^{1/2}$.
To make this point clear we also show the linear matter power spectrum for a larger DM mass.
Here we replace $m_{\chi} = 1$\, GeV by $100$\, GeV in Eqs.\,(\ref{eq:chipertdelta})--(\ref{eq:chipertpi}) while keeping $\gamma$ being intact.
The power spectra in the perfect and imperfect fluid approximations are concordant with each other in the plotted region ($k < 600 \, h/$Mpc).
This supports the above estimation of the effects of the higher order terms of the Boltzmann hierarchy.

\section{Effects of the free-streaming of hidden neutrinos}
\label{sec:freestream}
As discussed in Sec.\,\ref{sec:model}, in the phenomenological model considered in this paper, the scatterings between (hidden) neutrinos are controlled by the additional interaction such as the scalar mediator (see Eq.\,(\ref{eq:scalar})), while the pseudo scalar mediator (see Eq.\,(\ref{eq:pscalar})) is important for DM.
In Sec.\,\ref{sec:evol}, we assume that the scatterings between $\nu$'s are efficient and thus $\nu$ does not freely stream around the DM kinetic decoupling.
One may wonder how the free-streaming of hidden neutrinos impact the evolution of the DM perturbations and the resultant matter power spectrum.
To address this point, we consider the extreme case, i.e., the free-streaming neutrino limit, though in reality at least before the mediator becomes non-relativistic, the inverse decay of $\phi$ is efficient to prevent $\nu$ from streaming freely.

\begin{figure}[htb]
\begin{center}
\includegraphics[width=0.75\linewidth]{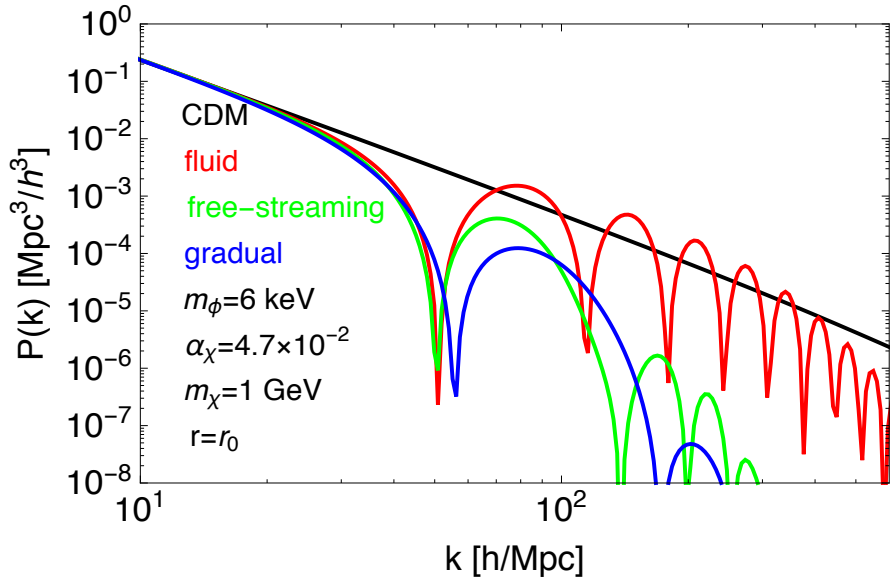}
\caption{Linear matter power spectrum in the free-streaming limit of $\nu$ is overplotted in Fig.\,\ref{fig:linear_pf_vb}.
The amplitudes of the oscillations are lower than those in the perfect fluid limit of $\nu$ due to the collisionless damping of the $\nu$ perturbations.
We also plot the linear matter power spectrum in the case of the gradual kinetic decoupling (scalar mediator case, $\gamma/H \propto T^{4}$) that is shown in Fig.\,2 in Ref.\,\cite{Binder:2016pnr}.
The amplitudes of the oscillations in this case are reduced when compared to those in the case of the sudden kinetic decoupling .
Note that $\nu$ is assumed to freely stream in the case of the gradual kinetic decoupling as well.
}
\label{fig:linear_pf_qr}
\end{center}
\end{figure}

In this limit, the evolution of the hidden neutrino perturbations are identical to those of the standard model neutrinos and thus not repeated here (see Eq.\,(49) of Ref.\,\cite{Ma:1995ey}).
We solve the co-evolution of the cosmological perturbations of DM, hidden neutrinos, and the other degrees of freedom numerically by suitably modifying \verb|CAMB|\,\cite{Lewis:1999bs}.
The resultant matter power spectrum is shown in Fig.\,\ref{fig:linear_pf_qr}.
It again shows the oscillating features, but the amplitudes of the peaks are suppressed when compared to those in the case of the perfect fluid that is the same as in Fig.\,\ref{fig:linear_pf_vb}.
This is because the free-streaming of $\nu$ transfers the initial powers of the density perturbation (monopole) and the bulk velocity potential (dipole) of $\nu$ to higher multipoles of the Boltzmann hierarchy (see, e.g., \cite{Ma:1995ey, Lesgourgues:2006nd}).
As we discussed in Sec.\,\ref{sec:evol}, the dark acoustic oscillation is driven by the bulk velocity potential of $\nu$.
It follows that the DM density perturbation and bulk velocity potential also lose their initial powers.
In the plot we show for reference the linear matter power spectrum in the case of the scalar mediator that is taken from Fig.\,2 in Ref.\,\cite{Binder:2016pnr}.
In this case $\nu$ is assumed to freely stream (i.e., free-steaming limit), while $\gamma/H \propto T^{4}$ and thus the kinetic decoupling is not instantaneous, but gradual.
This is why the ratio of the second peak to the first of the linear matter power spectrum in this case is larger than that in the case of the sudden kinetic decoupling.

\bibliographystyle{JHEP}
\bibliography{mdn}
  
\end{document}